%% file: main.tex
\documentclass[10pt,letterpaper]{article}
\usepackage[margin=0.78in]{geometry}
\usepackage{cite,url,courier}
\usepackage{amsmath,amssymb,amsthm,booktabs,tabularx,array,graphicx,xcolor,textcomp,microtype,enumitem,float}
\usepackage[hidelinks]{hyperref}
\hypersetup{pdftitle={Clock-Variable Discrete Simulated Bifurcation: A GPU Algorithm for Large-Scale Discrete-Phase Optimization},pdfauthor={Yaocheng Chen and Jiapei Zhuang},pdfsubject={Discrete-phase optimization with GPU-accelerated simulated bifurcation}}

\newcommand{\code}[1]{{\fontfamily{pcr}\fontseries{m}\selectfont#1}}

\newtheorem{proposition}{Proposition}
\graphicspath{{figures/}}
\title{Clock-Variable Discrete Simulated Bifurcation: A GPU Algorithm for Large-Scale Discrete-Phase Optimization}
\author{Yaocheng Chen\\
\small Instituto de Astronomia, Geof\'isica e Ci\^encias Atmosf\'ericas (IAG-USP),\\
\small Universidade de S\~ao Paulo, Rua do Mat\~ao 1226,\\
\small S\~ao Paulo, SP 05508-090, Brazil\\
\small \texttt{yaochengchen@usp.br}
\and
Jiapei Zhuang\\
\small Quantum Science Center of Guangdong-Hong Kong-Macao Greater Bay Area,\\
\small Shenzhen 518045, China\\
\small \texttt{jpzhuangphy@gmail.com}}
\date{2026}
\begin{document}
\maketitle
\begin{abstract}
We introduce clock-variable discrete simulated bifurcation (dSB), which represents each $M$-phase variable by a planar oscillator and evaluates the coupling at its nearest constellation point. For $M=2$, real couplings and real initial states, the update reduces exactly to dSB. We derive the local-field torque and conditions for improving cyclic-neighbor phase moves. Each step uses one real tensor-core matrix product and a fused update kernel, with CUDA Graph capture and multi-GPU row sharding. We evaluate the method on dense planted $\mathbb Z_M$ synchronization instances, including the pure clock spin glass, with $N=20{,}000$--$500{,}000$ and $M=4$--$64$. GPU methods share one engine, proxy-instance tuning and fp64 scoring on the same fp32-defined matrix. On the tested dense synchronization instances with $M\ge16$, clock-variable dSB achieves a higher objective than Gray-coded binary dSB on nearly every instance. At $N=20{,}000$, it also achieves a higher objective than the free-energy machine in 70 of 72 equal-wall-time comparisons. At $N=20{,}000$, $\lambda=1.1$ and $M=16,64$, at the tested budgets binary dSB requires about 3.5 times as long to match the 8.6~s clock-variable dSB result. Binary dSB performs best at $M=4$, while the free-energy machine remains competitive at $M=8$. One RTX PRO 6000 GPU sustains about 390~TFLOPS in FP16; eight GPUs process $N=400{,}000$ in FP16 and $N=500{,}000$ in INT8 at 85~ms per step. On modeled reconfigurable intelligent surface and phased-array problems with up to $128\times128$ elements, clock-variable dSB attains the best score on 20 of 24 surface instances, while the array instances yield near-ties across methods.
\end{abstract}
\noindent\textbf{Keywords:} Clock model, discrete phases, GPU computing, phased arrays, planar Potts model, reconfigurable intelligent surfaces, simulated bifurcation, tensor cores.

\noindent\textbf{Corresponding authors:} Yaocheng Chen (\url{yaochengchen@usp.br});\\ Jiapei Zhuang (\url{jpzhuangphy@gmail.com}).

\noindent\textbf{Funding:} Yaocheng Chen is supported by a FAPESP Postdoctoral Fellowship (Grant No. 25/25586-5).

\section{Introduction}
Scheduling, partitioning and routing problems can be written as Ising energies or quadratic functions of binary variables \cite{lucas,glover,boros}. Simulated annealing (SA) searches these objectives by flipping variables under a falling temperature \cite{kirkpatrick}. Simulated bifurcation (SB) evolves continuous oscillators whose bifurcation drives the variables toward two branches \cite{goto2019,goto2021}. Its simultaneous updates make the dominant operation a matrix product suited to GPUs.

A $K$-bit phase shifter in a phased array or reconfigurable intelligent surface (RIS) chooses one of $M=2^K$ phases per element \cite{wu2020}. The same discrete geometry appears in $M$-ary phase-shift keying, clock spin models \cite{honari2022} and $\mathbb Z_M$ synchronization \cite{perry2018}; angular synchronization uses continuous phases \cite{singer2011}. Clock interactions depend on the cosine of the phase difference, while standard Potts interactions use a Kronecker delta \cite{wu1982,honari2022}. For $M=2^K$, a binary encoding assigns $K$ spins to each phase variable. Under Gray coding, two one-bit directions follow the phase cycle and the others jump to nonadjacent phases. The encoded objective is multilinear in the bits, so the representation changes both the available moves and the coupling forces.

This paper replaces the binary encoding by a native representation in the dynamics. In clock-variable dSB each variable is a two-dimensional oscillator $z_j\in\mathbb C$; the bifurcation drives $|z_j|$ to a wall at one, and the coupling force is evaluated on the projection of $z_j$ onto the nearest of the $M$ constellation points, in the same way as discrete SB (dSB) evaluates it on $\operatorname{sgn}(x_j)$ \cite{goto2021}. For $M=2$ the method reduces exactly to dSB (Proposition~\ref{prop:m2}). For every $M$, each step evaluates one Hermitian matrix product against the projected constellation points. A stacked real--imaginary layout implements this product as one real GEMM, allowing us to reuse tensor cores, fused updates, graph capture and row sharding.

We address five questions using shared instances, a common GPU matrix-product engine, proxy-instance tuning and fp64-accumulated objective evaluation.
\begin{enumerate}[label=(Q\arabic*),leftmargin=*]
\item Does the clock representation outperform binary encodings of the same dSB dynamics, and for which $M$?
\item How does it compare at equal wall time with the following solvers: the free-energy machine (FEM) \cite{fem}, simulated annealing (SA) \cite{kirkpatrick,isakov2015}, a rank-one Burer--Monteiro relaxation \cite{burer2003,boumal2016}, the generalized power method (GPM) \cite{liu2017} and spectral rounding \cite{singer2011}?
\item How far can dense execution scale on one multi-GPU node, and which arithmetic precision is sufficient?
\item How does the solution quality compare on structured discrete-phase problems from RIS configuration and phased-array synthesis?
\item What does a better objective imply for the recovery of planted phases?
\end{enumerate}
The main results are:
\begin{itemize}
\item Clock-variable dSB uses projected complex positions and retains tangential momentum at a radial wall. We derive its reduction to binary dSB, local-field torque, neighboring-phase gain and within-sector stability (Section~\ref{sec:clock}). The C++/CUDA implementation shares its matrix-product engine with the GPU baselines (Section~\ref{sec:gpu}).
\item The representation comparison (Q1) favors binary dSB at $M=4$, gives mixed results at $M=8$, and favors clock-variable dSB at $M\ge16$. The advantage persists up to $N=400{,}000$ (Sections~\ref{sec:res-regime} and~\ref{sec:res-scaling}).
\item At equal wall time (Q2), clock-variable dSB wins 70 of 72 comparisons with FEM for $M\ge16$. Binary dSB requires about 3.5 times as long to match the 3000-step clock-variable dSB result at $\lambda=1.1$, $M=16,64$; 176-thread CPU SA does not reach that result within 1,100--2,600~s at $\lambda\le1.1$ (Sections~\ref{sec:res-time}--\ref{sec:res-sa}).
\item The engine sustains about 390~TFLOPS on one GPU in FP16 and runs dense instances of 400,000 (FP16) and 500,000 (INT8, FP8) variables on eight GPUs (Q3). The precision runs show no consistent FP16 quality loss relative to FP32; INT8 roughly halves the step time with larger objective gaps (Sections~\ref{sec:res-scaling} and~\ref{sec:res-precision}).
\item On RIS and phased-array problems with up to $128\times128$ elements (Q4), clock-variable dSB attains the best score on 20 of 24 RIS instances; the array instances yield near-ties. Designs are also evaluated with measured reflection states of a uniformly biased 4-bit RIS array \cite{zhang2025ris} (Sections~\ref{sec:app-method} and~\ref{sec:res-app}).
\item Near the spectral threshold, GPM recovers the planted phases more accurately despite a lower objective (Q5; Section~\ref{sec:res-overlap}).
\end{itemize}

The primary benchmark is the spiked $\mathbb Z_M$ synchronization model \cite{singer2011,perry2018}. Its signal strength $\lambda$ ranges from a pure clock spin glass at $\lambda=0$ to a signal-dominated problem; the spectral threshold is $\lambda=1$ \cite{feral2007}. A counter-based generator regenerates each coupling from its indices. RIS and array instances supply a second set of dense matrices derived from propagation and array-factor models. The implementation and result files are available in the public repository \cite{clockrepo}.

\section{Related work}\label{sec:related}
\subsection{Simulated bifurcation and Ising machines}
SB originated as a classical counterpart of adiabatic evolution in networks of nonlinear Kerr parametric oscillators \cite{goto2016,goto2019}. Its ballistic (bSB) and discrete (dSB) variants confine the positions by inelastic walls and update all variables simultaneously; in dSB the coupling acts on the signs of the positions only \cite{goto2021}, and it is this discrete-coupling principle that the present method generalizes from two to $M$ phase values. Later variants add thermal noise \cite{kanao2022}, handle cost functions of higher order \cite{kanao2023}, or combine the trajectories with tabu memory \cite{tabu2026}. Quantized SB discretizes the real position supplied to the matrix product into ternary or multi-level amplitudes, but its decision variable remains a binary Ising spin \cite{zhang2023}. On the large dense graphs studied in Ref.~\cite{zeng2024}, Zeng et al.\ report a shorter time to target for dSB than for the SB and simulated coherent-Ising-machine variants they evaluated.

Other physics-inspired solvers include coherent Ising machines \cite{inagaki2016}, digital annealers \cite{aramon2019} and oscillator networks \cite{mohseni2022}. Honari-Latifpour et al.\ evolve complex amplitudes or phase--angular-velocity pairs in a continuous clock potential \cite{honari2022}. Cheng and Lin formulate oscillator dynamics for categorical Potts variables \cite{cheng2025}, and Gonul et al.\ implement oscillator Ising/Potts machines on GPUs \cite{gonul2025}. The multi-direction-bifurcation patent of Yun et al.\ uses a complex variable with $q$ stable phases and continuous-state coupling \cite{yun2025}. Clock-variable dSB combines the pumped position--momentum update and wall of dSB with coupling evaluated on nearest-constellation projections. The binary backend derives from our GPU dSB implementation \cite{ourrepo}, whose dense matrix-product path is reused here.

\subsection{Multi-valued variables}
One-hot encodings use $M$ binary variables and a constraint per categorical variable \cite{lucas,glover}; compact encodings use $\lceil\log_2M\rceil$ bits and can produce higher-order interactions \cite{boros}. The free-energy machine (FEM) of Shen et al.\ assigns a softmax distribution to each categorical variable and minimizes an annealed variational free energy by gradient descent across many GPU replicas \cite{fem}. We implement this update in the common GPU engine.

\subsection{Phase synchronization}
Angular synchronization estimates phases from noisy pairwise measurements through eigenvector methods, semidefinite relaxations \cite{singer2011} or nonconvex optimization on the unit circle \cite{boumal2016,liu2017}. Liu et al.\ analyze the estimation error and convergence of the generalized power method \cite{liu2017}. Perry et al.\ develop approximate message passing for synchronization over compact groups, including $\mathbb Z_M$ \cite{perry2018}, and Javanmard et al.\ study phase transitions in semidefinite relaxations \cite{javanmard2016}. In the spiked Hermitian model used here, the top eigenvalue separates from the bulk at $\lambda=1$ \cite{feral2007}. Section~\ref{sec:res-overlap} compares objective values with phase recovery near this threshold.

\subsection{Discrete phases in applications}
Wu and Zhang study RIS beamforming with discrete phase shifts \cite{wu2020}. Our structured benchmarks use quadratic objectives from a Rician RIS channel and a phased-array model (Section~\ref{sec:app-method}). A 16-state reflection codebook digitized from the uniformly biased RIS array of Zhang et al.\ \cite{zhang2025ris} supplies measured amplitude and phase responses for post-design evaluation. A companion open-source implementation optimizes phased arrays and RISs directly over measured finite-state hardware tables through a factorized multilinear relaxation \cite{mprrepo}.

\subsection{Reduced precision on GPUs}
Markidis et al.\ measured tensor-core performance and rounding error for FP16 products with FP32 accumulation \cite{markidis2018}. Haidar et al.\ used these products in mixed-precision iterative refinement to obtain FP64-accuracy solutions \cite{haidar2018}. Here the product format controls the precision of the coupling force during the search. The final score uses fp64 accumulation on the shared fp32-defined matrix.

\section{Problem formulation}\label{sec:problem}
\subsection{Objective}
Let $\mathcal C_M=\{w^m e^{i\varphi_0}:m=0,\dots,M-1\}$ with $w=e^{2\pi i/M}$. Given a Hermitian $Q\in\mathbb C^{N\times N}$ with zero diagonal, we maximize
\begin{equation}
 f(\boldsymbol x)=\operatorname{Re}\,\boldsymbol x^{\mathsf H}Q\boldsymbol x=\sum_{j}\operatorname{Re}\big(\bar x_j h_j\big),\qquad \boldsymbol h=Q\boldsymbol x,\quad x_j\in\mathcal C_M,
 \label{eq:obj}
\end{equation}
and report $f/N$. Moving one variable from $x_j$ to $w^q e^{i\varphi_0}$ changes $f$ by $2\operatorname{Re}\big((w^qe^{i\varphi_0}-x_j)^*h_j\big)$, because $Q_{jj}=0$; a configuration is 1-opt when no such move is positive. For $M=2$ and real $Q$, \eqref{eq:obj} is an Ising problem.

\subsection{Planted model and regimes}
The instances are drawn from the spiked $\mathbb Z_M$ synchronization model
\begin{equation}
 Q=\frac{\lambda}{N}(\boldsymbol g\boldsymbol g^{\mathsf H}-I)+\frac{1}{\sqrt N}W,\qquad g_j=w^{m_j},
 \label{eq:model}
\end{equation}
with planted phases $m_j$ uniform on $\{0,\dots,M-1\}$ and $W$ a Gaussian Hermitian matrix with zero diagonal and off-diagonal entries $\mathcal{CN}(0,1)$. At $\lambda=0$, \eqref{eq:obj} is a clock spin glass with complex Gaussian couplings, the multi-phase analog of the Sherrington--Kirkpatrick model \cite{sk1975}. The top eigenvalue of $Q$ leaves the bulk edge at $2$ when $\lambda>1$ and tends to $\lambda+1/\lambda$ \cite{feral2007}; the measured values agree (for example, $2.007$ at $\lambda=1.1$, $N=20{,}000$, versus $2.009$). Asymptotically, below the threshold the top eigenvector has vanishing overlap with $\boldsymbol g$ \cite{perry2018}. The planted configuration has $f/N\approx\lambda$, well below the best objective found over the range studied here: at $\lambda=1.1$ and $M=16$, $f(\boldsymbol g)/N=1.10$, compared with $1.74$ for the best configurations found. Thus, $\lambda$ controls the strength of the planted direction in the noisy objective.

\subsection{Binary encodings}\label{sec:binary}
A binary encoding represents $x_j$ by $K=\log_2M$ spins $\boldsymbol s_j\in\{\pm1\}^K$ whose bit pattern $b$ (bit $k$ set when $s_{jk}<0$) selects a phase $x_j=w^{\pi(b)}e^{i\varphi_0}$, where $\pi$ is the identity (natural code) or the inverse binary-reflected Gray code. Under the Gray code a flip of the lowest bit always moves to a neighboring phase, while for $M\ge8$ higher-bit flips can make nonlocal moves with a shortest cyclic distance as large as $M/2-1$ phase steps; under the natural code a single flip can move by half a turn. For $M=4$ and $\varphi_0=0$, Gray coding gives $x_j=\tfrac{1-i}{2}s_{j0}+\tfrac{1+i}{2}s_{j1}$. Thus, \eqref{eq:obj} becomes an Ising problem on $2N$ spins. For $M\ge8$, the encoded objective is multilinear in each variable's $K$ spins. The binary dSB baseline (Section~\ref{sec:binary-dsb}) uses a force proportional to the exact gradient of the multilinear extension.

\begin{figure}[!t]
\centering
\includegraphics[width=0.98\textwidth]{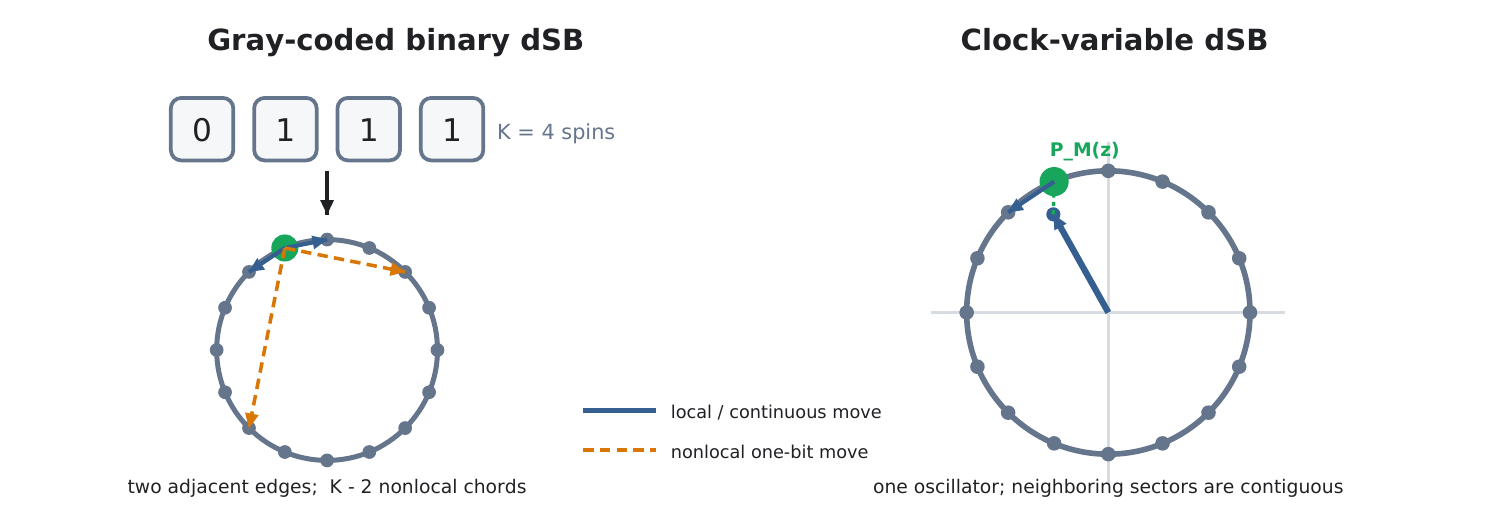}
\caption{Schematic of the representation geometry for $M=16$. A Gray-coded variable has two one-bit edges along the phase cycle and $K-2$ nonlocal chords. Clock-variable dSB projects the angle of a planar oscillator onto the nearest constellation point. Crossing one sector boundary changes the phase to a cyclic neighbor; larger angular steps can skip sectors.}
\label{fig:representation}
\end{figure}

\section{Clock-variable dSB}\label{sec:clock}
\subsection{Dynamics}
Each variable $j$ of each replica carries a position $z_j\in\mathbb C$ and a momentum $p_j\in\mathbb C$. With the discrete configuration $x_j=P_M(z_j)$, where $P_M(z)$ is the point of $\mathcal C_M$ nearest in argument to $z$, and $\boldsymbol h=Q\boldsymbol x$, one step is
\begin{align}
 p_j &\leftarrow p_j+\Delta t\,\big[-(a_0-a_t)\,z_j+c_0\,h_j\big],\label{eq:p}\\
 z_j &\leftarrow z_j+\Delta t\,a_0\,p_j,\label{eq:z}
\end{align}
followed by a wall. The pump rises as $a_t=a_0t/T$ over $T$ steps, $a_0=1$, and $c_0=\xi/(\sigma\sqrt N)$, where $\sigma$ is the root-mean-square off-diagonal $|Q_{jk}|$ and $\xi$ a dimensionless gain. Three walls were tested. The \emph{radial} wall, used by default, rescales $z_j$ to $|z_j|=1$ when it leaves the unit disk and removes only the radial component of the momentum, $p_j\leftarrow p_j-\operatorname{Re}(p_j\bar u)u$ with $u=z_j/|z_j|$, so that the tangential motion that rotates the phase survives the collision. The \emph{full} wall sets $p_j=0$. The \emph{square} wall clips the real and imaginary parts separately to $[-1,1]$ and resets the corresponding momentum components. Initial positions and momenta are uniform in a disk of radius $0.1$.

Equations~\eqref{eq:p}--\eqref{eq:z} extend the dSB update of Goto et al.\ \cite{goto2021} to a complex variable, replacing the sign with projection onto $\mathcal C_M$. The force $c_0h_j$ is proportional to the Wirtinger gradient of $z^{\mathsf H}Qz$, evaluated at $z=\boldsymbol x$. Constellation values are rounded or quantized in the selected product format. Final configurations are scored separately with fp64 accumulation.

\begin{table}[!t]
\centering\small
\caption{Gray-coded binary dSB and clock-variable dSB representations. Both use the same constellation, local-field product $\boldsymbol h=Q\boldsymbol x$ and pump schedule.}
\label{tab:representation}
\setlength{\tabcolsep}{5pt}
\begin{tabularx}{\textwidth}{p{0.14\textwidth}XX}
\toprule
Aspect & Gray-coded binary dSB & Clock-variable dSB \\
\midrule
State per variable & $K=\log_2M$ real oscillators, each with position and momentum: $2K$ real numbers & One planar oscillator with complex position $z_j$ and momentum $p_j$: four real numbers, independent of $M$ \\
Discrete phase & Signs of the $K$ positions form a code word, which is decoded to one of $M$ phases & The argument of $z_j$ is projected directly to the nearest point of $\mathcal C_M$ \\
Coupling force & One force per bit, proportional to the objective change under that bit flip, equivalently the multilinear-extension derivative at a vertex & The two-dimensional force $c_0h_j$ points along the complex local field evaluated at the projected configuration \\
Wall & Each real oscillator is clamped separately at $\pm1$ and its momentum is reset & $|z_j|$ is clamped at one; the radial wall removes radial momentum but retains tangential momentum \\
Phase motion & A one-bit flip follows one hypercube edge: two edges are cyclic neighbors and $K-2$ are nonlocal chords & Continuous angular motion crosses adjacent sectors when the step is smaller than one sector width \\
Per-step cost & The same dominant matrix product $Q\boldsymbol x$, plus $O(K)$ scalar state updates per variable & The same dominant matrix product $Q\boldsymbol x$, plus $O(1)$ scalar state updates per variable \\
\bottomrule
\end{tabularx}
\end{table}

\begin{proposition}\label{prop:m2}
Let $M=2$, $\varphi_0=0$, $Q$ real symmetric, and let all initial positions and momenta be real. Then clock-variable dSB with any of the three walls generates the same trajectory as dSB with the inelastic wall.
\end{proposition}
\begin{proof}
For $M=2$ and $\varphi_0=0$, $P_M(z)$ is $+1$ for $\operatorname{Re}z>0$ and $-1$ for $\operatorname{Re}z<0$, that is $\operatorname{sgn}(\operatorname{Re} z)$ up to the tie convention. If $z$ and $p$ are real, $\boldsymbol x$ and $\boldsymbol h=Q\boldsymbol x$ are real, so \eqref{eq:p}--\eqref{eq:z} keep both real and coincide with the dSB update. When $|z|>1$ the radial wall maps $z$ to $\operatorname{sgn}(z)$ and removes the radial momentum, which for real $p$ is all of it; the full and square walls do the same. By induction the trajectories agree.
\end{proof}

\subsection{Local objective geometry}
The coordinate-wise optimum follows directly from the local field.
\begin{proposition}\label{prop:best-response}
Fix all variables except $x_j$ and let $h_j=\sum_{k\ne j}Q_{jk}x_k$. If $h_j\ne0$, a best response is
\begin{equation}
 x_j^{\rm br}=P_M(h_j),
 \qquad |\arg(\bar x_j^{\rm br}h_j)|\le \frac{\pi}{M}.
 \label{eq:best-response}
\end{equation}
If $h_j=0$, every constellation point is a best response. A configuration is 1-opt if and only if each coordinate with $h_j\ne0$ is a nearest constellation point to its local field, with either choice allowed at a tie. A best response also satisfies
\begin{equation}
 \operatorname{Re}(\bar x_j^{\rm br}h_j)
 \ge |h_j|\cos(\pi/M).
 \label{eq:quant-bound}
\end{equation}
\end{proposition}
\begin{proof}
Because $Q_{jj}=0$, $h_j$ is independent of $x_j$, and the part of the objective that changes with $x_j=u$ is $2\operatorname{Re}(\bar u h_j)$. For $h_j=0$ this term vanishes for every $u$. Otherwise, maximizing it selects the constellation point with the smallest angular distance from $h_j$. That distance is at most half a sector, giving \eqref{eq:best-response} and \eqref{eq:quant-bound}. Applying the same condition at each coordinate gives the 1-opt criterion.
\end{proof}
The worst local-field quantization loss is $1-\cos(\pi/M)=O(M^{-2})$. Binary and clock representations encode the same discrete objective and search it through different state updates.

\subsection{Local-field torque and improving phase moves}
Let $p_j^+$ denote the momentum immediately after \eqref{eq:p}, before the position update and wall.
\begin{proposition}\label{prop:torque}
Assume $z_j\ne0$ and $h_j\ne0$, and write $z_j=r_je^{i\theta_j}$ and $h_j=|h_j|e^{i\psi_j}$. Define the angular momentum $L_j=\operatorname{Im}(\bar z_jp_j)$. Then the momentum update satisfies
\begin{equation}
 L_j^+-L_j
 =\Delta t\,c_0\operatorname{Im}(\bar z_jh_j)
 =\Delta t\,c_0r_j|h_j|\sin(\psi_j-\theta_j).
 \label{eq:torque}
\end{equation}
If a radial-wall collision maps a post-position state $\tilde z_j=\rho u$ and $\tilde p_j$ with $\rho>1$ and $|u|=1$ to
$z_j^{\rm w}=u$ and $p_j^{\rm w}=\tilde p_j-\operatorname{Re}(\tilde p_j\bar u)u$, then
\begin{equation}
 L_j^{\rm w}=\operatorname{Im}(\bar z_j^{\rm w}p_j^{\rm w})
 =\rho^{-1}\operatorname{Im}(\bar{\tilde z}_j\tilde p_j),
 \label{eq:wall-angular}
\end{equation}
so the wall preserves the sign of the angular momentum.

Now fix the other variables, let $x_j=e^{i\phi_j}\in\mathcal C_M$, set $\delta_j=\operatorname{Arg}(\bar x_jh_j)\in(-\pi,\pi]$, and let $\Delta\phi=2\pi/M$. For $0<|\delta_j|<\pi$, the cyclic neighbor in the direction of the local field is $x_j'=x_je^{i\operatorname{sgn}(\delta_j)\Delta\phi}$, and its exact objective change is
\begin{align}
 \Delta f_j
 &:=f(x_j',\boldsymbol x_{-j})-f(x_j,\boldsymbol x_{-j})\notag\\
 &=2|h_j|\big[\cos(|\delta_j|-\Delta\phi)-\cos|\delta_j|\big]\notag\\
 &=4|h_j|\sin(\pi/M)\sin(|\delta_j|-\pi/M).
 \label{eq:neighbor-gain}
\end{align}
Consequently, this neighboring move strictly improves the objective if and only if $|\delta_j|>\pi/M$, is neutral at equality, and decreases it if $|\delta_j|<\pi/M$.
\end{proposition}
\begin{proof}
The pump adds a real multiple of $z_j$ to $p_j$, so its contribution to $\operatorname{Im}(\bar z_jp_j)$ is zero. The coupling term gives \eqref{eq:torque}. At a radial collision, the subtracted momentum is a real multiple of $u$ and has zero angular momentum; rescaling $\tilde z_j=\rho u$ gives \eqref{eq:wall-angular}. Because $Q_{jj}=0$, changing only $x_j$ changes the objective by $2\operatorname{Re}[(\bar x_j'-\bar x_j)h_j]$. Substitution of $x_j'$ gives the first line of \eqref{eq:neighbor-gain}, and $\cos(a-b)-\cos a=2\sin(b/2)\sin(a-b/2)$ gives the second. For $0<\pi/M\le\pi/2$ and $0<|\delta_j|<\pi$, its sign is the sign of $|\delta_j|-\pi/M$.
\end{proof}
Equation~\eqref{eq:torque} shows that, for a nonzero, non-antipodal phase difference, the coupling injects angular momentum along the shortest angular direction from $z_j$ to its local field. Equation~\eqref{eq:neighbor-gain} gives the matching discrete statement: whenever the current phase lies outside the best-response sector, the cyclic-neighbor move in that direction improves the objective. With a frozen local field, repeated improving neighbor moves terminate at a best response. The radial wall retains the sign of this angular motion through a collision, whereas the full wall erases it by setting the momentum to zero. Inertia determines whether a step crosses a sector boundary, while simultaneous updates change the local fields.

\subsection{Piecewise dynamics}
Away from projection boundaries and walls, the discrete configuration and hence $h_j$ are constant. Freezing the pump at $a_t$ and writing $\kappa=a_0-a_t>0$, the affine update has fixed point $z_j^*=c_0h_j/\kappa$, $p_j^*=0$. For a fixed point inside the projection cell and wall, the local dynamics are as follows.
\begin{proposition}\label{prop:sector-stability}
Within one fixed projection cell, the linear part of one symplectic-Euler step about $(z_j^*,0)$ is
\begin{equation}
 \begin{bmatrix}z^+-z_j^*\\ p^+\end{bmatrix}
 =
 \begin{bmatrix}
 1-\Delta t^2a_0\kappa & \Delta t\,a_0\\
 -\Delta t\,\kappa & 1
 \end{bmatrix}
 \begin{bmatrix}z-z_j^*\\p\end{bmatrix}.
 \label{eq:sector-map}
\end{equation}
It is linearly stable for $0<\Delta t^2a_0\kappa<4$, but it is not asymptotically convergent.
\end{proposition}
\begin{proof}
The matrix in \eqref{eq:sector-map} has determinant one and trace $2-\Delta t^2a_0\kappa$. Its two eigenvalues are a complex-conjugate pair on the unit circle precisely when the stated inequality holds. Unit determinant precludes contraction.
\end{proof}
The pump, projection boundaries and wall collisions change this map during a run. The radial wall dissipates radial momentum while retaining tangential momentum, so sector crossings can continue after the pump reaches its final value. After the pump reaches $a_t=a_0$, a state with $z_j=x_j$, $p_j=0$ is absorbing under the radial wall if every local field has the alignment $h_j=\rho_jx_j$ with $\rho_j\ge0$. The wider best-response sector of Proposition~\ref{prop:best-response} permits 1-opt configurations that continue moving under the dynamics.

\subsection{Representation graph}
Figure~\ref{fig:representation} makes the distinction from binary dSB precise.
\begin{proposition}\label{prop:gray-graph}
For $M=2^K$ with $K\ge2$, binary-reflected Gray ordering embeds the phase cycle $C_M$ as a Hamiltonian cycle of the bit hypercube $Q_K$. Every encoded phase has $K$ one-bit neighbors, exactly two of which are adjacent on $C_M$; the remaining $K-2$ are nonlocal phase moves.
\end{proposition}
\begin{proof}
Consecutive words in the reflected Gray sequence, including the last and first, differ in one bit, so the $M$ cyclic phase edges are hypercube edges. A vertex of a cycle has two cyclic neighbors and a vertex of $Q_K$ has $K$ distinct one-bit neighbors, leaving $K-2$ edges outside the cycle.
\end{proof}
For $K=2$ ($M=4$), $Q_2=C_4$ and there are no chord directions; this is also the case in which the phase is linear in the two spins and the objective is an exact Ising model. At $K=3$ each state has one chord direction, while $K\ge4$ introduces at least two. The fraction of one-bit directions that follow a cyclic-neighbor edge is therefore exactly
\begin{equation}
 \frac{2}{K}=\frac{2}{\log_2M},
 \label{eq:gray-neighbor-fraction}
\end{equation}
which decreases from $1$ at $M=4$ to $1/3$ at $M=64$. In clock-variable dSB, an angular step smaller than $2\pi/M$ crosses at most one sector boundary. The local-field torque of Proposition~\ref{prop:torque} therefore acts through cyclic-neighbor moves at this step scale. The increasing number of Gray-code chords offers a geometric explanation for the representation dependence observed in Section~\ref{sec:res-regime}.

\subsection{Parameters}\label{sec:params}
The projection $P_M$ is piecewise constant and hence has no ordinary linearization at sector boundaries. We choose the time-step scale from the continuous surrogate obtained by replacing $P_M(z)$ with $z$. Along the most negative eigenmode at $t=0$, that surrogate has stiffness $k_0=a_0(a_0+c_0|\lambda_{\min}(Q)|)$. To avoid period-two oscillations near the symplectic-Euler stability limit, we set
\begin{equation}
 \Delta t=r\cdot\operatorname{clip}\!\left(\sqrt{2/k_0},\ 0.2,\ 2.5\right),
 \label{eq:dt}
\end{equation}
with $\lambda_{\min}(Q)$ estimated by Lanczos iteration on the stored matrix, and select the factor $r$ and the gain $\xi$ by the tuning procedure of Section~\ref{sec:tuning}. The selected gains lie between $0.3$ and $0.8$ and the factors between $0.8$ and $1.25$.

\subsection{Binary dSB baseline}\label{sec:binary-dsb}
The binary baseline integrates one dSB oscillator $(x_{jk},y_{jk})$ per spin with the inelastic wall. Let $F$ be the multilinear extension of $f$, and let $\Delta_{jk}=2\operatorname{Re}((x'_j-x_j)^*h_j)$ be the change of $f$ when bit $k$ of variable $j$ is flipped from the current pattern to give phase $x'_j$. At a vertex, $\partial F/\partial s_{jk}=-s_{jk}\Delta_{jk}/2$. The implemented force, $-\operatorname{sgn}(x_{jk})\Delta_{jk}/4$, is therefore one half of this exact derivative; the factor is absorbed into the tuned coupling gain $\xi$. It uses the same matrix product as clock-variable dSB, with $\boldsymbol h$ evaluated at the decoded configuration, so the two methods differ only in the representation of a variable and the update of its state.

\section{GPU implementation}\label{sec:gpu}
\subsection{Matrix product}
Write $Q'=\sqrt NQ=A+iB$, with $A$ symmetric and $B$ antisymmetric, and collect the configurations of $R$ replicas in $X=U+iV\in\mathbb C^{N\times R}$. The engine stores $[A;B]$ as one $2N\times N$ real matrix and evaluates
\begin{equation}
 \begin{bmatrix} C_A\\ C_B\end{bmatrix}=\begin{bmatrix}A\\ B\end{bmatrix}\begin{bmatrix}U & V\end{bmatrix},\qquad
 \begin{aligned}
 \operatorname{Re}H&=(C_{AU}-C_{BV})/\sqrt N,\\
 \operatorname{Im}H&=(C_{BU}+C_{AV})/\sqrt N,
 \end{aligned}
 \label{eq:gemm}
\end{equation}
with one \code{cublasGemmEx} call of shape $(2N\times N)(N\times2R)$, that is $8N^2R$ floating-point operations per step. The scale $\sqrt N$ keeps the stored entries of order one at every $N$, which avoids FP16 subnormals at $N=500{,}000$. Supported formats are FP16 and BF16 with FP32 accumulation, TF32, FP32, INT8 with INT32 accumulation (couplings clipped at five standard deviations, configurations scaled by 127) and FP8 (e4m3, through cuBLASLt). The INT8 and FP8 kernels require the configuration in replica-major order, so a transpose precedes each of their products.

\subsection{Fused update and graph capture}
One fused kernel per step, with one thread per (variable, replica), reads the product, accumulates $\operatorname{Re}(\bar x_jh_j)$ into per-block energy partials, commits the best configuration of the previous step for the replicas that improved, advances the method's state, projects it, and writes the next configuration directly in the input format of the next product. A small kernel of $R$ threads sums the partials in a fixed order, so the energies and the best-so-far configurations are bitwise reproducible; it also records checkpoints of the best energy. On one GPU the whole run, $T+1$ products with their update and reduction kernels, is captured into one CUDA Graph \cite{cuda} and launched once. All methods of Table~\ref{tab:methods} that run on the GPU use this structure; they differ only in the elementwise update and in the per-element state, which is four floats for clock-variable dSB and $2K$ floats for binary dSB.

\subsection{Multiple GPUs}
For several GPUs the rows of $[A;B]$ and of the state are divided into contiguous shards. Each GPU multiplies its rows against the full configuration and updates its variables; it then copies its rows of the next configuration and its $2R$ energy partials to every peer with \code{cudaMemcpyPeerAsync}, and peers wait on per-buffer events before their next product. The eight GPUs used here are connected through PCIe without NVLink, and the copies are not overlapped with the products in the present implementation.

The present solver runs as an in-process application on the GPU server. A companion binary-dSB architecture handles request reception, computation and replies on the GPU and streams dense coupling tiles for models larger than device memory \cite{gpunetio2026}; its implementation and experimental data are public \cite{gpunetiorepo}. Extending that network-facing path to complex constellation states would enable a remotely accessed clock-variable dSB accelerator.

\subsection{Common objective evaluation and refinement}
For the random benchmark, every entry $Q'_{jk}$ is a pure function of the seed and the unordered pair $\{j,k\}$, drawn with the counter-based Philox4x32-10 generator \cite{salmon2011}; the noise uses one draw for $(j,k)$ and $(k,j)$, so $A$ is exactly symmetric and $B$ exactly antisymmetric in fp32, and the matrix is generated in place on each GPU, directly in the storage format, identically for any number of shards. We take this deterministically rounded fp32 matrix as the benchmark instance. Because any entry can be regenerated, the engine recomputes the objective of a configuration from those fp32 entries with fp64 accumulation, without storing a second matrix.

For the structured benchmarks, a \code{--qfile} path supplies an explicit zero-diagonal Hermitian complex64 matrix. The same file is read by the GPU product, the fp64-accumulated scorer and the 1-opt refinement; a separate proxy-file pattern supplies disjoint tuning instances of the same problem class. After either type of run, the 16 replicas with the highest tracked energy are rescored, and the best 8 receive exhaustive best-improvement 1-opt refinement using the same finite-precision entries and the move gain of Section~\ref{sec:problem}; the refinement time is part of the reported time of every method. Every method is therefore ranked by fp64 accumulation on the same finite-precision instance. BM and FEM evolve continuous relaxed states; their rounded configurations are evaluated at checkpoints and at the final step, whereas the discrete methods track configurations at every step.

\subsection{Capacity}
In FP16 the stored matrix takes $4N^2$ bytes: $N\approx140{,}000$ fits one 96~GB GPU, and eight GPUs hold $N=400{,}000$ (80~GB per GPU). In INT8 or FP8 it takes $2N^2$ bytes, and $N=500{,}000$ occupies 62.5~GB per GPU.

\section{Experimental methodology}\label{sec:method}
\subsection{Hardware}
Table~\ref{tab:config} lists the configuration. GPU runs use one or eight NVIDIA RTX PRO 6000 Blackwell Server Edition GPUs; the CPU baseline uses the host processors of the same server.

\begin{table}[!t]\centering\small
\caption{Experimental configuration.}
\label{tab:config}
\begin{tabularx}{\columnwidth}{lX}\toprule
Item & Configuration \\
\midrule
GPUs & 8$\times$ NVIDIA RTX PRO 6000 Blackwell Server Edition (97{,}887~MiB each, 600~W cap), PCIe, no NVLink \\
CPU & 2$\times$ Intel Xeon 6747P (48 cores, 96 threads per socket, up to 3.9~GHz, 576~MiB L3 in total), 4 NUMA nodes \\
System & Linux 5.14.0 (Enterprise Linux 9.6), NVIDIA driver 610.57.04 (CUDA 13.3); cuBLAS and cuBLASLt; C++17; OpenMP for the CPU baseline \\
Replicas & 256 per run (128 for the eight-GPU runs); 176 for SA \\
Precision & FP16 storage with FP32 accumulation unless stated \\
\bottomrule
\end{tabularx}
\end{table}

\subsection{Instances and campaigns}
All instances have $\varphi_0=0$. The main campaigns use $N=20{,}000$, $M\in\{4,8,16,32,64\}$ and $\lambda\in\{0,0.8,1.0,1.1,\allowbreak 1.2,1.35,1.5,2.0\}$ with 16 instances (seeds 0--15) per cell; a step-budget sweep at $T\in\{300,1000,3000,10{,}000,30{,}000\}$; single-GPU runs at $N\in\{40,80,120,140\}\times10^3$; eight-GPU runs at $N\in\{200,300,400\}\times10^3$ (FP16) and $500\times10^3$ (INT8 and FP8); a precision ablation; and reference runs with $T=30{,}000$ steps for $M\in\{16,64\}$. The baseline campaign repeats $N=20{,}000$, $M\in\{8,16,32,64\}$, $\lambda\in\{0,1.1,1.5\}$ on seeds 0--7 with all methods of Table~\ref{tab:methods}; a seed denotes the same instance in every campaign. In total 6,698 result records on 734 instances enter the analysis.

\subsection{Structured instances}\label{sec:app-method}
We construct two families of discrete phase-shifter problems. For $X_j\in\mathcal C_M$, both minimize
\begin{equation}
 y(\boldsymbol X)=\sum_{\ell} a_\ell\left|d_\ell+(B\boldsymbol X)_\ell\right|^2,
 \label{eq:app-objective}
\end{equation}
where negative $a_\ell$ reward power at intended users or in the main beam and positive $a_\ell$ penalize leakage or sidelobe power. Expanding $-y$ gives $J=-B^{\mathsf H}\operatorname{diag}(\boldsymbol a)B$, whose diagonal is constant for unit-modulus variables and is removed, and a linear term with $\boldsymbol b=-B^{\mathsf H}\operatorname{diag}(\boldsymbol a)\boldsymbol d$, which is homogenized by one auxiliary clock variable,
\begin{equation}
 \widetilde Q=\begin{bmatrix}J&\boldsymbol b\\\boldsymbol b^{\mathsf H}&0\end{bmatrix};
 \label{eq:app-augment}
\end{equation}
the returned configuration is rotated so that the auxiliary variable equals one. Each matrix is written once in complex64 with unit off-diagonal RMS and read by the GPU and CPU solvers alike; physical scores are independently recomputed from $B$, $\boldsymbol d$ and the returned phases using complex64 matrix products, followed by fp64 power evaluation and accumulation. The \emph{RIS} family uses a half-wavelength square array, one far-field base-station direction, eight intended users and four protected points over four subcarriers (10\% bandwidth), Rician factor 10 and a weak direct path, so $\operatorname{rank}(J)\le48$ before removing the diagonal. The \emph{array} family rewards 64 samples inside the main lobe of a random direction and penalizes $2N$ samples outside 1.5 times the first-null radius, so $J$ has full rank. Both use $64\times64$ (eight instances) and $128\times128$ elements (four instances; $N=4096$ or $16{,}384$, plus one for RIS) at $M=16$ and $64$; tuning uses separate $64\times64$ instances, the GPU methods run $T=3000$ steps with 256 replicas in FP16, and SA runs on 22 threads per group. For $M=16$ the returned phases are also re-evaluated with the 16 reflection coefficients measured at 3.6~GHz on a uniformly biased $20\times10$ 4-bit RIS array, digitized from Fig.~6(d,e) of \cite{zhang2025ris}, using the best of 16 cyclic index offsets. The measured array responses are applied as a per-element codebook in the modeled propagation geometry.

\begin{table}[!t]\centering\small
\caption{Methods and tuning grids. GPU methods share one matrix product per step, fp64-accumulated scoring and 1-opt refinement. In the random baseline campaign, FEM and binary dSB use the second proxy tuning, with the wider grids in brackets and four proxy seeds.}
\label{tab:methods}
\begin{tabularx}{\textwidth}{lXX}\toprule
Method & Update per step & Tuned parameters \\
\midrule
clock-dSB (radial, full, square) & Eqs.~\eqref{eq:p}--\eqref{eq:z} with the wall of Section~\ref{sec:clock} & $\xi\in\{0.3,\dots,1.6\}$ (8 values), $r\in\{0.8,1,1.25,1.5,1.8\}$ \\
binary dSB (Gray, natural) & dSB per spin, multilinear-extension force (Section~\ref{sec:binary-dsb}) & $\xi\in\{1,\dots,8\}$ (7 values), $r\in\{0.8,\dots,1.8\}$ [up to 2.7] \\
FEM \cite{fem} & softmax marginals over $M$ phases; free energy $-f(\boldsymbol m)-H/\beta$, $\boldsymbol m$ the mean phase; Adam \cite{kingma2015} on the logits; $\beta$ geometric from $\beta_{\min}$ to $\beta_{\max}=b(M/2\pi)^2$; argmax rounding & $\beta_{\min}\in\{0.1,0.2,0.5,1\}$, $b\in\{1,2,5,10\}$ [2--50], learning rate $\{0.1,0.3,1\}$ [up to 2] \\
BM (rank one) \cite{burer2003,boumal2016} & $z_j\leftarrow P_{\mathrm{circle}}(z_j+\eta h_j+s_tn_j)$, $s_t=s_0(1-t/T)$, $n_j\sim\mathcal{CN}(0,1)$; nearest-point rounding & $s_0\in\{0,\dots,0.8\}$, $\eta\in\{0.1,\dots,1000\}$ \\
GPM \cite{liu2017} & $x_j\leftarrow P_M(h_j+\mu x_j)$ from the rotated top eigenvector with noise; restart when stalled or after a fixed period & $\mu\in\{0,\dots,2\}$, period $\in\{50,\dots,1000\}$ \\
spectral \cite{singer2011} & top eigenvector rounded to $\mathcal C_M$, best of 64 global rotations & -- \\
SA (CPU) \cite{kirkpatrick} & heat-bath sweeps, geometric temperature $T_0\to T_1$, zero-temperature quench to 1-opt; 176 independent replicas on 176 threads & $T_0\in\{0.5,1,2\}$, $T_1=c\min(1,(2\pi/M)^2)$, $c\in\{0.03,\dots,1\}$ \\
\bottomrule
\end{tabularx}
\end{table}

\subsection{Tuning}\label{sec:tuning}
For each method and $(M,\lambda,T,\text{precision})$, all grid points run as replicas of one batch on proxy instances with $N=8192$. Proxy seeds start at $10^6$ and are disjoint from the evaluation seeds. The grid point with the highest mean final objective supplies the parameters for every evaluation replica. The random baseline campaign reports FEM and binary dSB from the second tuning throughout, using the wider grids in Table~\ref{tab:methods} and four proxy seeds. In this FEM tuning, $\beta_{\min}=0.5$ and the learning rate were interior to the grid; $b$ reached 50 in two of 18 settings. The structured benchmarks use two separate $64\times64$ proxy instances from each family and reuse the selected parameters at both sizes. Tuning and Lanczos estimation of $\lambda_{\min}$ are excluded from the reported run time. GPM and spectral rounding include Lanczos time because the top eigenvector supplies their initial configuration.

\subsection{Timing}
The reported time of a GPU method is the wall time of the dynamics, measured around the launch of the captured graph (or the eager loop on several GPUs) after a warm-up of the same kernels, plus the exhaustive 1-opt refinement, plus the Lanczos time for GPM and spectral rounding. Instance generation, warm-up, graph capture and the fp64 rescoring of candidates are excluded and recorded separately. For SA the reported time is the wall time of all 176 replicas running in parallel, including the computation of the initial local fields and the final quench, with instance generation and tuning excluded.

\subsection{Metrics}
For every random instance the reference $f_{\mathrm{ref}}$ is the best fp64-accumulated objective found by any method in any campaign, including the long reference runs, and the gap of a result is $100(f_{\mathrm{ref}}-f)/f_{\mathrm{ref}}$ in percent. Tables report the mean gap over instances and paired counts, the number of instances on which one method finds a strictly higher objective than another. Since $f$ scales with $N$, a gap of 0.01\% at $N=20{,}000$ corresponds to an objective difference of about $3.5$ in units of $f$. For the time-to-target comparison the target on an instance is the objective of clock-variable dSB with $T=3000$, and a method reaches it at a budget when its result is within a relative $10^{-5}$ of it on at least half of the instances. For an application instance, where smaller $y$ is better and $y$ may be negative because desired power is rewarded, the reference $y_{\mathrm{ref}}$ is the smallest independently recomputed physical score among all methods and the reported gap is $100(y-y_{\mathrm{ref}})/|y_{\mathrm{ref}}|$. For application best counts and paired comparisons, scores within a relative $10^{-7}$ are treated as tied. We also report intended-region and leakage/sidelobe powers, wall time, paired best counts, and, for $M=16$, the corresponding gap after replacing ideal phases by the measured reflection codebook.

\section{Results}\label{sec:results}
\subsection{Validation}\label{sec:res-validation}
A self-test compares the GPU engine with a host reference built from the same elementwise code. The checks cover matrix Hermiticity and zero diagonal, GEMM layout and format-specific error, shard invariance, all eight GPU dynamics methods, graph and eager execution, reproducibility, fp64 rescoring and 1-opt refinement. All checks pass on the RTX PRO 6000. From the same initial state, the clock-variable dSB trajectory agrees with the original complex128 PyTorch prototype on every phase, and the best energies agree to a relative $3.6\times10^{-8}$. On a small external-matrix CPU test, converting the engine objective back to 
\eqref{eq:app-objective} agrees with direct evaluation of the returned configuration to relative error $7.6\times10^{-8}$; over all 432 records of the GPU and CPU application campaign the largest such discrepancy is $2.5\times10^{-7}$.

\subsection{Representation and regime}\label{sec:res-regime}
\begin{figure}[!t]\centering
\includegraphics[width=\textwidth]{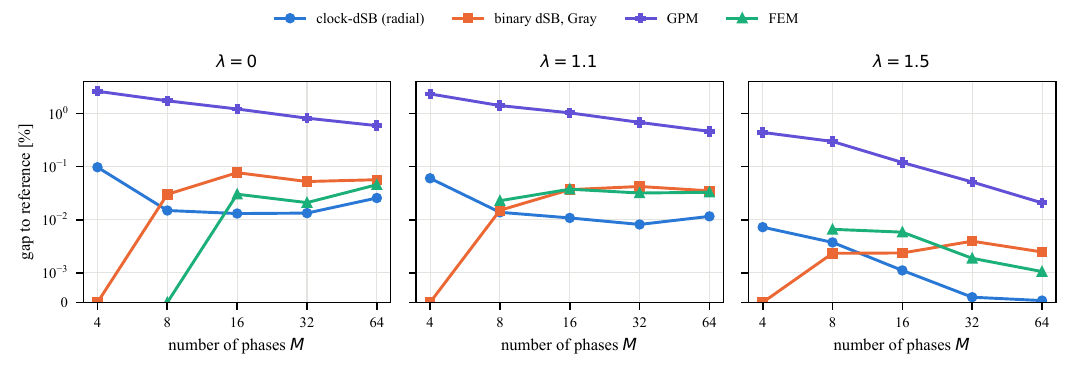}
\caption{Mean gap to the reference against the number of phases $M$ ($N=20{,}000$, $T=3000$, FP16; 16 instances per point for clock-variable dSB, binary dSB and GPM, 8 instances for FEM using its second proxy tuning). The vertical axis is logarithmic above $10^{-3}\%$ and linear below; a point at zero attains the reference on every instance.}
\label{fig:regime}
\end{figure}

Table~\ref{tab:night} and Figure~\ref{fig:regime} compare the variable representations at $N=20{,}000$ and $T=3000$ on 16 instances per cell. The ordering depends on $M$ and is stable in $\lambda$.

For $M=4$ the Gray-coded binary dSB attains the reference on every instance, and clock-variable dSB lies 0.007--0.10\% below it: here the Gray code is an exact decomposition into $2N$ Ising spins (Section~\ref{sec:binary}), and the binary dynamics solve that Ising problem directly. At $M=8$ the two are close, with clock-variable dSB ahead on 13, 10 and 4 of 16 instances at $\lambda=0$, $1.1$ and $1.5$. From $M=16$ on clock-variable dSB has the higher objective on 15 or 16 of 16 instances in every cell, and its gap is two to six times smaller than that of the Gray code at $\lambda\le1.1$ and more than an order of magnitude smaller at $\lambda=1.5$ and $M\ge32$.

The variants of each representation are not interchangeable. The radial wall has a smaller gap than the full wall in every cell of Table~\ref{tab:night}, by a factor of 1.7 to 7 at $\lambda\le1.1$, showing the benefit of retaining tangential momentum. The square wall treats the real and imaginary parts as independent dSB coordinates and leaves raw gaps of 5--10\% before refinement. Among binary codes, the natural code trails the Gray code, with raw gaps of 1.6--8.7\%; its single-bit moves can span half a turn. The refinement also shows how close each method ends to a local optimum: the median number of 1-opt moves on the best replicas is 2 for clock-variable dSB, 789 for binary dSB, $8.9\times10^4$ for GPM and $1.8\times10^5$ for spectral rounding. GPM and spectral rounding remain 0.02--2.6\% below the reference after refinement.

\begin{table}[!t]\centering\small
\caption{Mean gap to the reference [\%] by representation ($N=20{,}000$, $T=3000$, FP16, 16 instances per row, after 1-opt refinement). The last column counts the instances on which clock-variable dSB (radial) finds a strictly higher objective than Gray-coded binary dSB.}
\label{tab:night}
{\begin{tabular}{rr ccc cc cc c}\toprule
 & & \multicolumn{3}{c}{clock-dSB} & \multicolumn{2}{c}{binary dSB} & & & clock $>$ \\
\cmidrule(lr){3-5}\cmidrule(lr){6-7}
$M$ & $\lambda$ & radial & full & square & Gray & natural & GPM & spectral & Gray \\
\midrule
\input{tables/tab_night.tex} 
\bottomrule
\end{tabular}}
\end{table}

\subsection{Comparison with the baselines}\label{sec:res-baselines}
Table~\ref{tab:main} compares the baselines on seeds 0--7, using the second proxy tuning for FEM and binary dSB. At $M\ge16$ and $\lambda\ge1.1$, clock-variable dSB finds a higher objective than FEM on all 48 instances and than binary dSB on 46 of 48 at 3000 steps. On the pure spin glass, clock-variable dSB wins 5 of 8 comparisons with FEM at $M=16$ and 3 of 8 at both $M=32$ and $64$. FEM has the lowest mean gap at $M=8$, $\lambda=0$, attaining the reference on all eight instances. BM, GPM and spectral rounding stay 0.002--1.8\% below the reference.

FEM takes 1.22, 1.47, 1.99 and 3.08 times the wall time of clock-variable dSB at $M=8$, 16, 32 and 64 for the same step budget. Its softmax and Adam updates act on $M$ logits per variable and replica; the matrix-product dimensions remain fixed. Section~\ref{sec:res-equal} compares the methods at equal wall time.

\begin{table}[!t]\centering\small
\caption{Mean gap to the reference [\%] at $T=3000$ ($N=20{,}000$, FP16, seeds 0--7). FEM and binary dSB use the second proxy tuning. The paired columns count instances with a higher clock-variable dSB objective at equal steps and, in the final column, at the wall time of FEM (Table~\ref{tab:eqtime}).}
\label{tab:main}
{\begin{tabular}{rr cccccc ccc}\toprule
 & & \multicolumn{6}{c}{gap to reference [\%]} & \multicolumn{3}{c}{clock-dSB higher} \\
\cmidrule(lr){3-8}\cmidrule(lr){9-11}
$M$ & $\lambda$ & clock-dSB & binary dSB & FEM & BM & GPM & spectral & vs binary & vs FEM & vs FEM, equal time \\
\midrule
\input{tables/tab_main.tex} 
\bottomrule
\end{tabular}}
\end{table}

\subsection{Quality against wall time}\label{sec:res-time}
\begin{figure}[!t]\centering
\includegraphics[width=\textwidth]{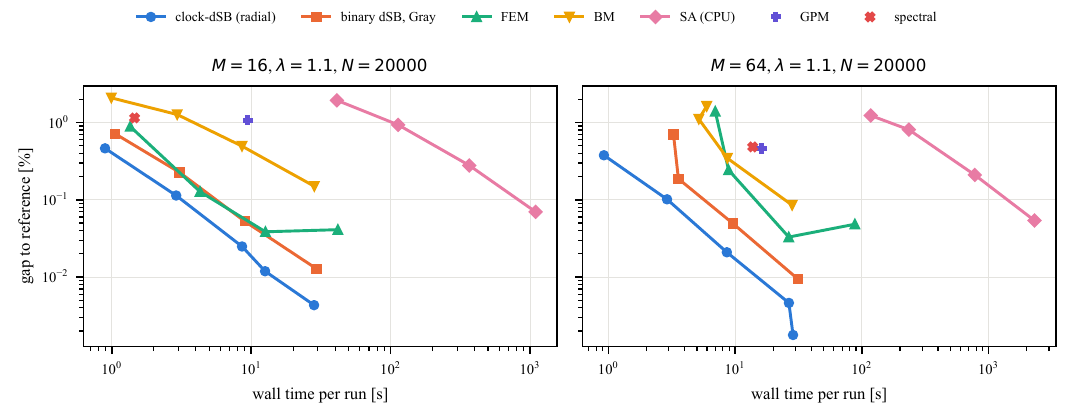}
\caption{Mean gap to the reference against wall time per run ($N=20{,}000$, $\lambda=1.1$, seeds 0--3). Points along a line are step budgets $T\in\{300,1000,3000,10{,}000\}$ for the GPU methods (plus the equal-time budget of Table~\ref{tab:eqtime} for clock-variable dSB) and sweep budgets $\{100,300,1000,3000\}$ for SA on 176 CPU threads; GPM and spectral rounding are single points at $T=3000$. GPU times include the 1-opt refinement.}
\label{fig:time}
\end{figure}

Figure~\ref{fig:time} compares objective quality with wall time at $\lambda=1.1$ for $M=16$ and $64$. For $M=16$, the clock-variable dSB gap falls from 0.46\% at 300 steps (0.9~s) to 0.025\% at 3000 steps (8.6~s) and 0.0043\% at 10,000 steps (28~s). For $M=64$, the last two gaps are 0.021\% and 0.0018\%. Binary dSB first reaches the 3000-step clock target at the 10,000-step budget: 29.5~s for $M=16$ and 31.3~s for $M=64$, or 3.4 and 3.6 times the clock run time. FEM improves up to 3000 steps, then its mean gap rises from 0.039\% to 0.041\% for $M=16$ and from 0.033\% to 0.048\% for $M=64$ at 10,000 steps. It reaches neither target at the tested budgets. The plateau appears with both tuning grids under the geometric inverse-temperature schedule. BM improves more slowly; GPM and spectral rounding have gaps one to two orders of magnitude larger than clock-variable dSB at comparable times.

\subsection{Equal wall time against FEM}\label{sec:res-equal}
For the equal-wall-time comparison in Table~\ref{tab:eqtime}, we run clock-variable dSB for 3700, 4400, 6000 and 9300 steps at $M=8$, 16, 32 and 64, respectively. These budgets match the measured wall time of 3000 FEM steps to within 1\%. For $M\ge16$ clock-variable dSB then finds a higher objective on 70 of the 72 instances, including 22 of 24 on the pure spin glass, where FEM led at equal steps. For $M=8$ FEM remains better on the spin glass (8 of 8) and on 6 of 8 instances at $\lambda=1.5$, and clock-variable dSB is better on 7 of 8 at $\lambda=1.1$.

\begin{table}[!t]\centering\small
\caption{Equal wall time: clock-variable dSB with $T_{\mathrm{c}}$ steps against FEM with 3000 steps and its second proxy tuning ($N=20{,}000$, seeds 0--7). Times in seconds, gaps in percent; the final column counts higher clock-variable dSB objectives.}
\label{tab:eqtime}
\setlength{\tabcolsep}{3.2pt}
\begin{tabular}{rrrrrrrr}\toprule
$M$ & $\lambda$ & $T_{\mathrm c}$ & $t_{\mathrm{clock}}$ & $t_{\mathrm{FEM}}$ & gap$_{\mathrm{clock}}$ & gap$_{\mathrm{FEM}}$ & wins \\
\midrule
\input{tables/tab_eqtime.tex} 
\bottomrule
\end{tabular}
\end{table}

\subsection{CPU simulated annealing}\label{sec:res-sa}
Table~\ref{tab:sa} compares the CPU heat-bath SA with clock-variable dSB on one GPU. SA keeps exact local fields and moves one variable at a time; each accepted move costs one pass over a row of $Q$, so on 176 threads a 100-sweep run on $N=20{,}000$ variables takes 41~s for $M=16$ and 118~s for $M=64$. At $\lambda\le1.1$, 3000 sweeps (1,100--2,600~s) leave SA at gaps of 0.05--0.11\%, two to three times the gap of clock-variable dSB at 8.6~s, and SA reaches the target on none of these instances. At $\lambda=1.5$, SA reaches the target on three of four instances at 3000 sweeps for $M=16$ (121 times the GPU time) and on all four at 1000 sweeps for $M=64$ (70 times).

\begin{table}[!t]\centering\small
\caption{CPU heat-bath SA (176 threads) against clock-variable dSB ($T=3000$, one GPU), $N=20{,}000$, seeds 0--3. Times in seconds, gaps in percent; ``reach'' counts instances on which SA comes within a relative $10^{-5}$ of clock-variable dSB.}
\label{tab:sa}
\setlength{\tabcolsep}{3.2pt}
\begin{tabular}{rrrrrrrr}\toprule
 & & \multicolumn{2}{c}{clock-dSB} & \multicolumn{4}{c}{SA} \\
\cmidrule(lr){3-4}\cmidrule(lr){5-8}
$M$ & $\lambda$ & $t$ & gap & sweeps & $t$ & gap & reach \\
\midrule
\input{tables/tab_sa.tex} 
\bottomrule
\end{tabular}
\end{table}

\subsection{Scaling}\label{sec:res-scaling}
\begin{figure}[!t]\centering
\includegraphics[width=\textwidth]{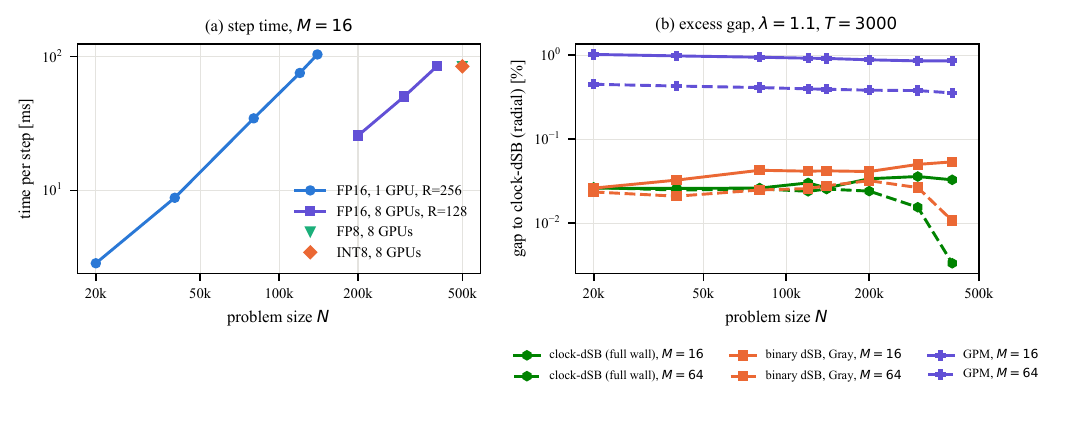}
\caption{(a) Time per step of clock-variable dSB at $M=16$, $\lambda=1.1$ on one GPU ($R=256$) and on eight GPUs ($R=128$); the INT8 and FP8 points at $N=500{,}000$ coincide. (b) Gap of the other dSB variants and of GPM to clock-variable dSB (radial) on the same instance at $\lambda=1.1$, $T=3000$; solid lines $M=16$, dashed $M=64$.}
\label{fig:scaling}
\end{figure}

Table~\ref{tab:scaling} and Figure~\ref{fig:scaling}(a) give the step time of clock-variable dSB. On one GPU it grows as $N^2$ from $N=40{,}000$ on, with a sustained 370--390~TFLOPS on the product counted as $8N^2R$ operations; at $N=20{,}000$ the fixed cost of the update kernels lowers the rate to 288~TFLOPS. On eight GPUs with $R=128$, $N=200{,}000$, 300,000 and 400,000 run at 25.6, 50.2 and 84.5~ms per step, 1.6--1.9~PFLOPS in total or 200--240~TFLOPS per GPU, which is 52--63\% of the single-GPU rate; the difference comes from exchanging configuration rows over PCIe between products. INT8 and FP8 at $N=500{,}000$ run at the same 84.5~ms per step as FP16 at 400,000.

Figure~\ref{fig:scaling}(b) shows a persistent advantage over the other dSB variants as $N$ grows. At $\lambda=1.1$ and $T=3000$, Gray-coded binary dSB remains 0.01--0.05\% below clock-variable dSB from $N=20{,}000$ to 400,000 for both $M=16$ and $64$. Among the methods in the scaling campaign, the radial wall returns the highest objective on every instance with $N\ge40{,}000$ and $M=16$. Clock-variable dSB also exceeds the second-tuned FEM result on both $N=80{,}000$ baseline instances at each of $M=16$ and $64$.

\begin{table}[!t]\centering\small
\caption{Step time and throughput of clock-variable dSB ($M=16$, $\lambda=1.1$). Throughput counts $8N^2R$ arithmetic operations per step; TOPS denotes $10^{12}$ operations per second, floating-point for FP formats and integer for INT8.}
\label{tab:scaling}
\setlength{\tabcolsep}{3.5pt}
\begin{tabular}{rrlrrrr}\toprule
$N$ & GPUs & format & $R$ & ms/step & TOPS & per GPU \\
\midrule
\input{tables/tab_scaling.tex} 
\bottomrule
\end{tabular}
\end{table}

\subsection{Arithmetic precision}\label{sec:res-precision}
Table~\ref{tab:precision} compares product formats on shared instances. FP16, BF16 and TF32 show no consistent quality loss relative to FP32 across the listed seeds. At $N=20{,}000$, FP16 and BF16 run 4.0--4.1 times faster than FP32, and TF32 runs 2.6 times faster. At $N=80{,}000$, the speedups are 5.7--5.9 for FP16 and BF16; the FP32 product uses CUDA cores. INT8 roughly halves the FP16 step time, reaching 7.9 and 11.5 times the FP32 speed at the two sizes, with larger gaps in every listed setting. FP8 has similar speed and the largest gaps. We use FP16 by default and INT8 for the largest instances.

\begin{table}[!t]\centering\small
\caption{Precision of the stored product for clock-variable dSB ($M=16$, $T=3000$; seeds 0--3 at $N=20{,}000$, seed 0 at $N=80{,}000$). Gaps in percent, computed by fp64 accumulation on the common fp32-defined instance.}
\label{tab:precision}
\setlength{\tabcolsep}{4pt}
\begin{tabular}{rrlrrrr}\toprule
$N$ & $\lambda$ & format & $n$ & ms/step & vs FP32 & gap \\
\midrule
\input{tables/tab_precision.tex} 
\bottomrule
\end{tabular}
\end{table}

\begin{table}[!t]\centering\scriptsize
\caption{RIS instances: mean gap [\%] to the best physical score found on each instance ($T=3000$, FP16; SA uses the sweep budget with the lowest mean gap in each setting among 100, 300 and 1000 sweeps; 0.0000 is below $5\times10^{-5}$). On the array instances all gaps are below $5\times10^{-5}$.}
\label{tab:app-results}
\setlength{\tabcolsep}{2.2pt}
\begin{tabular}{lrrrrrrrr}
\toprule
Array & $M$ & clock & binary & FEM & BM & GPM & spectral & SA \\
\midrule
$64^2$ & 16 & 0.0000 & 0.0001 & 0.0002 & 0.0005 & 0.0006 & 0.0100 & 0.0001 \\
$64^2$ & 64 & 0.0000 & 0.0001 & 0.0000 & 0.0000 & 0.0004 & 0.0004 & 0.0000 \\
$128^2$ & 16 & 0.0068 & 0.0387 & 0.0062 & 0.0392 & 1.2749 & 1.2766 & 0.0002 \\
$128^2$ & 64 & 0.0000 & 0.0307 & 0.0000 & 0.0307 & 1.2687 & 1.2686 & 0.0000 \\
\bottomrule
\end{tabular}
\end{table}

\subsection{Objective and recovery}\label{sec:res-overlap}
\begin{figure}[!t]\centering
\includegraphics[width=\columnwidth]{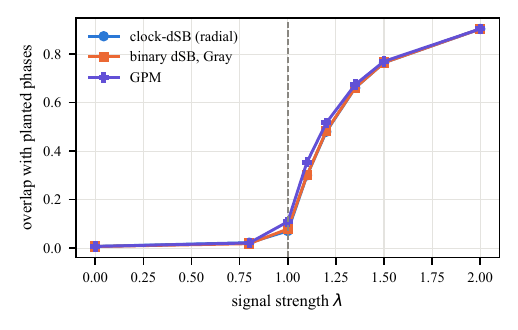}
\caption{Overlap of the best configuration with the planted phases against $\lambda$ ($N=20{,}000$, $M=16$, $T=3000$, 16 instances per point). The dashed line marks the spectral threshold $\lambda=1$.}
\label{fig:overlap}
\end{figure}

Figure~\ref{fig:overlap} plots the overlap $|\tfrac1N\sum_j\bar g_jx_j|$ with the planted phases at $M=16$. The three methods have small overlap below $\lambda=1$ and increasing overlap above it. Between $\lambda=1.1$ and $1.35$, GPM has higher overlap than clock-variable dSB despite a lower objective: at $\lambda=1.1$, the overlaps are 0.354 and 0.300, while the GPM objective is about 1\% lower. Near the threshold, maximizing the likelihood can improve the fit to noise without improving phase recovery. Approximate message passing addresses recovery through a model of the phase posterior \cite{perry2018}.

\subsection{Structured instances}\label{sec:res-app}
The physical score recomputed from each saved configuration agrees with the converted solver objective to a relative $2.5\times10^{-7}$ over all 432 records. The full-rank array instances do not separate the methods: every method, including spectral rounding, attains the same score on at least 22 of 24 instances, because the main-beam term dominates the matrix and its leading eigenvector is already almost optimal. Table~\ref{tab:app-results} gives the RIS results. Clock-variable dSB attains the best score on 20 of 24 instances and matches or exceeds BM, GPM and spectral rounding on all 24. Against binary dSB, FEM and SA with 1000 sweeps, its better/tied/worse counts are 15/8/1, 13/9/2 and 9/12/3. Clock-variable dSB, binary dSB, FEM, BM and SA have mean gaps below 0.04\% in every RIS setting, while GPM and spectral rounding lose about 1.3\% at $128\times128$. The cost per step differs more than the result: at $128\times128$, clock-variable dSB takes 1.66--1.68~ms, binary dSB 1.75--1.86~ms and FEM 2.76--6.74~ms ($M=16$--64). The measured reflection codebook introduces a loss of about 2.3~dB in the objective magnitude, $10\log_{10}(|y_{\rm meas}|/|y_{\rm ideal}|)\approx-2.3$~dB, for every method. Most of the loss comes from the codebook's mean reflected power ($-2.2$~dB). The ranking among clock-variable dSB, binary dSB, FEM, BM and SA changes at the 0.05\% level, while their margin over GPM and spectral rounding persists.

\section{Discussion}\label{sec:discussion}
\subsection{Why the clock representation helps}
The coupling adds angular momentum toward the local-field phase (Proposition~\ref{prop:torque}). When the current symbol lies outside the field's best-response sector, moving to the cyclic neighbor in that direction increases the objective. The radial wall retains tangential momentum through collisions, allowing continued sector crossings. Gray coding also provides both cyclic-neighbor edges, but these account for only $2/\log_2M$ of its one-bit directions; the remaining edges are nonlocal chords (Proposition~\ref{prop:gray-graph}). The chord count grows from zero at $M=4$ to four at $M=64$, matching the observed change in representation performance. Clock-variable dSB also needs a median of only two subsequent 1-opt moves in the representation campaign.

\subsection{FEM}
FEM is competitive at equal steps, particularly on the spin glass and at $M=8$. Its softmax, gradient and Adam updates scale with the $M$ logits per variable and replica. Clock-variable dSB uses four real state values per variable, and its step cost is dominated by the matrix product. At $M=64$, FEM takes about three times longer for the same step budget. At $\lambda=1.1$, extending FEM from 3000 to 10,000 steps under the geometric inverse-temperature schedule leaves the mean gap unchanged or larger.

\subsection{Throughput and scaling}
Dense clock-variable dSB is compute-bound on one GPU, and its step time follows $N^2R$ at 370--390~TFLOPS in FP16. The multi-GPU path loses up to half of that rate in the exchange of configuration rows over PCIe. Splitting each product into the block of local columns and the blocks received from peers would let the transfers overlap with computation, and exchanging INT8 configurations would halve the transferred bytes. Capacity is set primarily by the stored matrix: INT8 or FP8 requires $2N^2$ bytes for the real and imaginary parts, in addition to state and workspace. Sparse, structured or streamed couplings would extend execution beyond the dense in-memory limit.

\section{Conclusion}\label{sec:conclusion}
Clock-variable dSB extends the projected-coupling update of dSB to $M$ discrete phases with one planar oscillator per variable. Its local-field torque directs angular motion, and the radial wall retains the tangential momentum needed to cross phase sectors. A single real tensor-core matrix product per step supports dense execution up to 400,000 variables in FP16 and 500,000 in INT8 on eight GPUs.

On the planted synchronization benchmark, the clock representation performs best for $M\ge16$, binary dSB for $M=4$, and FEM remains competitive at $M=8$. At $N=20{,}000$ and $\lambda=1.1$, binary dSB needs about 3.5 times as long to match the clock-variable dSB result at $M=16,64$. Clock-variable dSB wins 70 of 72 equal-time comparisons with FEM for $M\ge16$. CPU SA on 176 threads does not match the 8.6~s clock-variable dSB result within 1,100--2,600~s at $\lambda\le1.1$. Near the spectral threshold, GPM has higher phase-recovery overlap despite a lower objective.

On the modeled RIS and array problems, the same engine handles up to $128\times128$ elements. Clock-variable dSB attains the best score on 20 of 24 RIS instances, while the arrays yield near-ties across methods. The measured reflection codebook introduces a similar reflection loss for every method. Larger application matrices could be evaluated through their channel or array structure instead of dense storage; overlapping peer transfers with products would improve the present multi-GPU path.

\section*{Code and Data Availability}
The clock-variable implementation, external-matrix generator, run instructions, analysis scripts and result files are available at \url{https://github.com/yaochengchen/clock-dsb} \cite{clockrepo}. The random and structured instances are defined by their seeds and regenerated by the code; the digitized 4-bit codebook records its published figure and frequency source. The binary-dSB GPU code from which its execution path was derived is available separately \cite{ourrepo}; the companion remote-request and streaming implementation and its experimental data are also public \cite{gpunetio2026,gpunetiorepo}. The measured-codebook optimization software used in related work is available at \url{https://github.com/yaochengchen/mpr-codebook-optimization} \cite{mprrepo}.
\section*{Author Contributions}
Yaocheng Chen proposed the research idea, developed the software, and performed the computational experiments. Jiapei Zhuang contributed to scientific discussions and the analysis of the final results. Yaocheng Chen and Jiapei Zhuang jointly wrote and revised the manuscript.

\section*{Acknowledgment}
Y. Chen acknowledges support from FAPESP (grants 25/25586-5 and 2021/02120-0). Generative AI tools (OpenAI ChatGPT, Anthropic Claude and Google Gemini) assisted with software development; ChatGPT was also used for language editing. The authors reviewed and verified all AI-assisted output and are responsible for the manuscript and software.

\section*{Conflict of Interest}
The authors declare no conflicts of interest.

\phantomsection
\section*{Author Biographies}
\paragraph{Yaocheng Chen.}
Yaocheng Chen received the B.S. degree in physics and electronics from Central South University in 2015, and the M.S. and Ph.D. degrees in physics from National Taiwan University in 2017 and 2025, respectively. He is a FAPESP postdoctoral fellow at IAG-USP, Universidade de S\~ao Paulo, Brazil. Previously, he was an algorithm scientist at Jingdong, China; a project researcher and visiting scholar at the Quantum Science Center of Guangdong-Hong Kong-Macao Greater Bay Area; and a KASI postdoctoral fellow at the Korea Astronomy and Space Science Institute (KASI), Republic of Korea. His interests include high-energy astroparticle physics, extragalactic astrophysics, GPU computing and FPGA systems. He is a member of the ASTRI Mini-Array and TAROGE collaborations, the CTAO Consortium and the Space Array team. His previous collaborations include ANITA/PUEO, ARA and BURSTT; he was also a member of the ALMA Total Power GPU Spectrometer (TPGS) team. ORCID: \url{https://orcid.org/0000-0002-8967-4911}.

\phantomsection
\paragraph{Jiapei Zhuang.}
Jiapei Zhuang received the Ph.D. degree from National Tsing Hua University in 2022. He was a postdoctoral fellow at the Southern University of Science and Technology. In 2024, he joined the Quantum Science Center of Guangdong-Hong Kong-Macao Greater Bay Area as an assistant researcher. His research interests include quantum algorithms, quantum computing, cold-atom quantum simulation, physics-inspired computing, and AI for science. ORCID: \url{https://orcid.org/0000-0002-9140-5139}.

\end{document}

%% file: tables/tab_night.tex
4 & 0 & 0.0974 & 0.2437 & 2.1468 & 0.0000 & 2.2356 & 2.6201 & 2.6436 & 0/16 \\
4 & 1.1 & 0.0603 & 0.1763 & 2.0806 & 0.0000 & 2.2552 & 2.3318 & 2.3960 & 0/16 \\
4 & 1.5 & 0.0073 & 0.0263 & 0.5353 & 0.0000 & 0.4505 & 0.4408 & 0.4716 & 0/16 \\
\midrule
8 & 0 & 0.0149 & 0.1037 & 0.9556 & 0.0301 & 2.0304 & 1.7333 & 1.7644 & 13/16 \\
8 & 1.1 & 0.0138 & 0.0625 & 0.9219 & 0.0151 & 1.7604 & 1.4103 & 1.4483 & 10/16 \\
8 & 1.5 & 0.0038 & 0.0116 & 0.3773 & 0.0023 & 0.3393 & 0.2976 & 0.3157 & 4/16 \\
\midrule
16 & 0 & 0.0131 & 0.0590 & 0.4055 & 0.0770 & 1.5031 & 1.2143 & 1.2320 & 16/16 \\
16 & 1.1 & 0.0108 & 0.0368 & 0.3824 & 0.0369 & 1.4259 & 1.0276 & 1.0618 & 16/16 \\
16 & 1.5 & 0.0011 & 0.0022 & 0.1243 & 0.0024 & 0.1401 & 0.1197 & 0.1258 & 16/16 \\
\midrule
32 & 0 & 0.0134 & 0.0349 & 0.2311 & 0.0525 & 1.2553 & 0.8165 & 0.8400 & 15/16 \\
32 & 1.1 & 0.0082 & 0.0267 & 0.2192 & 0.0423 & 1.0726 & 0.6801 & 0.7021 & 16/16 \\
32 & 1.5 & 0.0002 & 0.0006 & 0.0548 & 0.0040 & 0.0616 & 0.0515 & 0.0537 & 16/16 \\
\midrule
64 & 0 & 0.0257 & 0.0450 & 0.1771 & 0.0567 & 0.9335 & 0.5906 & 0.6213 & 15/16 \\
64 & 1.1 & 0.0116 & 0.0381 & 0.1536 & 0.0351 & 0.8254 & 0.4604 & 0.4856 & 15/16 \\
64 & 1.5 & 0.0001 & 0.0002 & 0.0236 & 0.0025 & 0.0237 & 0.0208 & 0.0257 & 16/16 \\

%% file: tables/tab_main.tex
8 & 0 & 0.0263 & 0.0508 & 0.0000 & 0.7054 & 1.7700 & 1.8160 & 8/8 & 0/8 & 0/8 \\
8 & 1.1 & 0.0108 & 0.0212 & 0.0229 & 0.5964 & 1.4043 & 1.4625 & 7/8 & 7/8 & 7/8 \\
8 & 1.5 & 0.0061 & 0.0000 & 0.0066 & 0.1851 & 0.2915 & 0.3310 & 0/8 & 3/8 & 2/8 \\
\midrule
16 & 0 & 0.0262 & 0.0785 & 0.0303 & 0.5007 & 1.2471 & 1.2603 & 8/8 & 5/8 & 7/8 \\
16 & 1.1 & 0.0217 & 0.0521 & 0.0376 & 0.4517 & 1.0598 & 1.0862 & 8/8 & 8/8 & 8/8 \\
16 & 1.5 & 0.0016 & 0.0023 & 0.0058 & 0.0440 & 0.1223 & 0.1231 & 6/8 & 8/8 & 8/8 \\
\midrule
32 & 0 & 0.0245 & 0.0745 & 0.0210 & 0.4420 & 0.8502 & 0.8832 & 8/8 & 3/8 & 7/8 \\
32 & 1.1 & 0.0163 & 0.0483 & 0.0319 & 0.3794 & 0.6794 & 0.7074 & 8/8 & 8/8 & 8/8 \\
32 & 1.5 & 0.0005 & 0.0031 & 0.0019 & 0.0101 & 0.0504 & 0.0525 & 8/8 & 8/8 & 8/8 \\
\midrule
64 & 0 & 0.0480 & 0.0763 & 0.0456 & 0.4492 & 0.6444 & 0.6675 & 8/8 & 3/8 & 8/8 \\
64 & 1.1 & 0.0207 & 0.0474 & 0.0329 & 0.3780 & 0.4826 & 0.5041 & 8/8 & 8/8 & 8/8 \\
64 & 1.5 & 0.0002 & 0.0023 & 0.0011 & 0.0025 & 0.0194 & 0.0260 & 8/8 & 8/8 & 8/8 \\

%% file: tables/tab_eqtime.tex
8 & 0 & 3700 & 10.5 & 10.4 & 0.0215 & 0.0000 & 0/8 \\
8 & 1.1 & 3700 & 10.5 & 10.4 & 0.0067 & 0.0229 & 7/8 \\
8 & 1.5 & 3700 & 10.5 & 10.4 & 0.0061 & 0.0066 & 2/8 \\
\midrule
16 & 0 & 4400 & 12.6 & 12.6 & 0.0161 & 0.0303 & 7/8 \\
16 & 1.1 & 4400 & 12.6 & 12.6 & 0.0095 & 0.0376 & 8/8 \\
16 & 1.5 & 4400 & 12.6 & 12.6 & 0.0011 & 0.0058 & 8/8 \\
\midrule
32 & 0 & 6000 & 17.2 & 17.1 & 0.0002 & 0.0210 & 7/8 \\
32 & 1.1 & 6000 & 17.2 & 17.1 & 0.0000 & 0.0319 & 8/8 \\
32 & 1.5 & 6000 & 17.2 & 17.1 & 0.0000 & 0.0019 & 8/8 \\
\midrule
64 & 0 & 9300 & 26.5 & 26.6 & 0.0126 & 0.0456 & 8/8 \\
64 & 1.1 & 9300 & 26.7 & 26.5 & 0.0034 & 0.0329 & 8/8 \\
64 & 1.5 & 9300 & 26.7 & 26.4 & 0.0000 & 0.0011 & 8/8 \\

%% file: tables/tab_sa.tex
16 & 0 & 8.6 & 0.0361 & 100 & 62 & 1.4090 & 0/4 \\
 & & & & 300 & 176 & 0.6159 & 0/4 \\
 & & & & 1000 & 570 & 0.2501 & 0/4 \\
 & & & & 3000 & 1716 & 0.1058 & 0/4 \\
\midrule
16 & 1.1 & 8.6 & 0.0249 & 100 & 41 & 1.9424 & 0/4 \\
 & & & & 300 & 113 & 0.9342 & 0/4 \\
 & & & & 1000 & 366 & 0.2791 & 0/4 \\
 & & & & 3000 & 1096 & 0.0699 & 0/4 \\
\midrule
16 & 1.5 & 8.6 & 0.0018 & 100 & 41 & 0.0389 & 0/4 \\
 & & & & 300 & 108 & 0.0153 & 0/4 \\
 & & & & 1000 & 351 & 0.0057 & 0/4 \\
 & & & & 3000 & 1039 & 0.0021 & 3/4 \\
\midrule
64 & 0 & 8.6 & 0.0564 & 100 & 122 & 1.0803 & 0/4 \\
 & & & & 300 & 263 & 0.7280 & 0/4 \\
 & & & & 1000 & 863 & 0.2998 & 0/4 \\
 & & & & 3000 & 2584 & 0.1096 & 0/4 \\
\midrule
64 & 1.1 & 8.6 & 0.0209 & 100 & 118 & 1.2344 & 0/4 \\
 & & & & 300 & 235 & 0.8094 & 0/4 \\
 & & & & 1000 & 781 & 0.2097 & 0/4 \\
 & & & & 3000 & 2307 & 0.0540 & 0/4 \\
\midrule
64 & 1.5 & 8.6 & 0.0002 & 100 & 70 & 0.0114 & 0/4 \\
 & & & & 300 & 185 & 0.0013 & 2/4 \\
 & & & & 1000 & 606 & 0.0006 & 4/4 \\

%% file: tables/tab_scaling.tex
20{,}000 & 1 & FP16 & 256 & 2.85 & 288 & 288 \\
40{,}000 & 1 & FP16 & 256 & 8.80 & 372 & 372 \\
80{,}000 & 1 & FP16 & 256 & 34.50 & 380 & 380 \\
120{,}000 & 1 & FP16 & 256 & 75.36 & 391 & 391 \\
140{,}000 & 1 & FP16 & 256 & 103.62 & 387 & 387 \\
200{,}000 & 8 & FP16 & 128 & 25.64 & 1598 & 200 \\
300{,}000 & 8 & FP16 & 128 & 50.20 & 1836 & 230 \\
400{,}000 & 8 & FP16 & 128 & 84.51 & 1939 & 242 \\
500{,}000 & 8 & FP8 & 128 & 84.56 & 3027 & 378 \\
500{,}000 & 8 & INT8 & 128 & 84.51 & 3029 & 379 \\

%% file: tables/tab_precision.tex
20000 & 1.2 & FP32 & 4 & 11.44 & 1.0 & 0.0092 \\
20000 & 1.2 & TF32 & 4 & 4.43 & 2.6 & 0.0081 \\
20000 & 1.2 & BF16 & 4 & 2.80 & 4.1 & 0.0063 \\
20000 & 1.2 & FP16 & 4 & 2.86 & 4.0 & 0.0058 \\
20000 & 1.2 & INT8 & 4 & 1.46 & 7.9 & 0.0178 \\
20000 & 1.2 & FP8 & 4 & 1.40 & 8.1 & 0.0376 \\
20000 & 1.5 & FP32 & 4 & 11.46 & 1.0 & 0.0016 \\
20000 & 1.5 & TF32 & 4 & 4.41 & 2.6 & 0.0016 \\
20000 & 1.5 & BF16 & 4 & 2.79 & 4.1 & 0.0012 \\
20000 & 1.5 & FP16 & 4 & 2.86 & 4.0 & 0.0026 \\
20000 & 1.5 & INT8 & 4 & 1.45 & 7.9 & 0.0065 \\
20000 & 1.5 & FP8 & 4 & 1.41 & 8.1 & 0.0240 \\
80000 & 1.5 & FP32 & 1 & 198.98 & 1.0 & 0.0002 \\
80000 & 1.5 & TF32 & 1 & 72.06 & 2.8 & 0.0000 \\
80000 & 1.5 & BF16 & 1 & 33.62 & 5.9 & 0.0005 \\
80000 & 1.5 & FP16 & 2 & 34.62 & 5.7 & 0.0016 \\
80000 & 1.5 & INT8 & 1 & 17.28 & 11.5 & 0.0057 \\
80000 & 1.5 & FP8 & 1 & 17.96 & 11.1 & 0.0215 \\